# Inside a Life-Threatening Crowd: Analysis of the Love Parade Disaster from the Perspective of Eyewitnesses

Anna Sieben[1,2] and Armin Seyfried[2,3]

[1] School of Humanities and Social Sciences, University of St. Gallen, Switzerland

[2] Institute for Advanced Simulation, Forschungszentrum Jülich, Jülich, Germany

[3] Faculty of Architecture and Civil Engineering, University of Wuppertal, Wuppertal, Germany

anna.sieben@unisg.ch, a.seyfried@fz-juelich.de

**Abstract**

During the Love Parade disaster in 2010 in Duisburg, Germany, twenty-one visitors lost their lives and more than five hundred were injured in a very dense crowd on the route to and from the festival area. Approximately nine hundred visitors who had been among this crowd were subsequently interviewed by police officers as eyewitnesses. This paper analyzes a random sample of 136 of these witness statements, focusing on how those present perceived the crowd, how they behaved, how they experienced the event emotionally, what happened to their bodies, and which collective dynamics they described. This approach provides a perspective from within crowd dynamics which are usually observed from a top-view perspective. Almost all the attendees became strongly focused on the staircase and the pole—the only visible ways out of the crowd. In some cases, they tried to reach these destinations by all means and at the expense of others. But at the same time, helping behavior is the behavior most frequently mentioned. Although witnesses described feelings of intense fear, they reject the idea of mass panic. As the most dangerous dynamics, a combination of falls (often after people had fainted) and transversal waves (which made people fall on top of each other) could be reconstructed. When people fall in a tightly packed crowd, a hole can form which pulls in more people due to the pressure on those standing around the edge of the hole, thus creating a pile of wedged bodies.

# 1. Introduction

Many projects and papers in the field of pedestrian dynamics take crowd disasters as their starting point. They aim thereby to understand crowd dynamics and the risks involved and to use this knowledge to make crowds safer. This includes scientific analysis of the London bombing (Drury et al., 2009), crowd disasters in Mekka (Helbing et al., 2007), or the disaster at "The Who" concert (Johnson, 1987), just to name a few examples. This paper contributes to this literature by reconstructing the dynamics that led to the crowd disaster in Duisburg, Germany, in 2010.

The term "crowd disaster", however, covers accidents with very different causes and courses of events. In order to specify our research approach different forms of accidents are distinguished. Some disasters are caused by external hazards like fires or terrorist attacks, whereas others are a product of the management of the crowd and the resulting dynamics within the crowd (e.g. miscalculation, incorrect estimation of the size of the crowd, problematic spatial situations or misleading communications). This paper focuses on this second case only and analyzes the Love Parade disaster in Germany in which an extremely dense crowd had fatal consequences in 2010 (Gerlach, 2021). Additional crowd disasters without external hazards but with dangerous internal dynamics are, for example,

- the Khodynka Tragedy at the coronation of the Czar Nicholas II in 1896 during which a distribution of gifts leads to a crush (Khodynka, Wikipedia, 2023),.
- the snowboard event at Bergisel Stadium in Austria in 1999 during which crowding occurred at a bottleneck in front of a temporary exit (Bergisel, Waldau, 2002),
- a fireworks display at a beach in Akashi, Japan, in 2001, where a congestion with standstill occurred on a pedestrian bridge with two-directional pedestrian flows (Tsuji, 2003), and
- the lantern festival at the Mihong Bridge in China in 2004, where again a bidirectional flow on a bridge led to a standstill (Zhen et al., 2008).

More events have been described in Rogsch et al. (2010), Illiyas et al. (2013), Feliciani et al. (2021) and on Keith Still's website (Still, 2023).

In this category, different dangerous dynamics within crowds can be distinguished. Without claim to completeness, we differentiate between:

a) Crowd rush: Low or medium density, high speed (running people), triggered, among other things, by (rumors of) external hazards, but also in situations where large crowds are moving quickly, such as, in Minsk in 1999, where a crowd rushed from an outdoor music festival to a metro station to escape a thunderstorm (Minsk, 1999). For crowd rush, alternatively the terms mass panic and stampede have been and are still used, although they have been criticized as inappropriate (see for an overview Lügering et al., 2023).

- b) Crowd shoving: Very high density, low speed, intentional or unintentional pushing. Instead of crowd shoving, this dynamic can be called "crowd crush". Within crowd shoving, the following collective phenomena are described as particularly risky:
  - Blockages due to clogging in front of bottlenecks in unidirectional (Dieckmann, 1911; Garcimartín et al., 2016; Müller, 1981; Muir et al., 1996; Sime, 1980) or multidirectional streams (Tsuji, 2003; Zhen et al., 2008).
  - Sudden collective movements and density waves, (Adrian et al., 2020; Bottinelli et al., 2016; Garcimartín et al., 2018, Oasis concert) also called turbulences (Helbing et al., 2007).
  - People falling over each other building a pile (Dieckmann, 1911; Helbing and Mukerji, 2012; Johnson, 1987; Schneider, 2001).

As the detailed expert report by Gerlach (see Section 2) and scientific reconstructions (Helbing and Mukerji, 2012) have shown, crowd shoving occurred at the Love Parade. We summarize the events that led to this extremely dense crowd in the second part of this paper.

The dangerous phenomena in dense crowds listed above have been empirically studied: The first studies of blockages at pedestrian bottlenecks were related to fire safety within theaters and are documented in Dieckman (1911). More references to studies on bottleneck flow and clogging can be found in Boltes et al. (2018) or Zuriguel (2014). Additional investigations have addressed the question of the influence of motivation and spatial layout of the bottleneck on the density in front of the bottleneck and how density and pressure due to pushing affect the formation of blockages (Adrian et al., 2020; Garcimartín et al., 2016; Muir et al., 1996; Sieben et al., 2017; Yanagisava et al., 2009). Predtechenskii and Milinskii (1978) were the first to report on collective movement in high density crowds, describing how the movement of the crowd is "fused" at high density. Helbing et al. studied a crowd disaster that occurred in Mecca in 2006/2007 and called this phenomenon "crowd turbulence" to address the unordered character of the collective movement (Helbing et al., 2007). A similar approach can be found in Feliciani and Nishinari (2018). Sudden collective movements transversal to the intended moving direction are reported in experiments with pushing crowds at bottlenecks (Adrian et al., 2020; Garcimartín et al., 2018). Historical references to the problem of people falling over each other in dense crowds can be found in the description of a battle of knights in medieval times (Azincourt). Recent scientific investigations of situations connected with the risk of people falling on top of each other have been documented in Wang et al. (2018, 2020) and Li et al. (2021). This research focuses on how people lose their balance—see the review of Winter (1959)—under conditions where others' presence restrict individuals' range of actions to recover from stumbling. While it is certain that these phenomena are associated with life-threatening risks, one can only conjecture how they or their combination leads to injuries or deaths of individuals. Although one known cause of death is asphyxiation (e.g., Nolan et al., 2021) many questions remain unanswered, including, for example, the following: Can the pressure of a pushing crowd on the body of a person standing in the crowd lead to suffocation? What are the influences of hard edges on the body? Is there a correlation between the condition of the ground and the risk of tripping or the trigger of people falling on top of each other? One reason why these questions are still open is a methodological one: Most of the previously mentioned studies are characterized by the fact that the perspective for their observations concerning the crowd is from above or from the side. Here, two methodological limitations should be noted: Actual cameras used for making videos for scientific analysis of large crowds (e.g., as in Bottinelli, 2018; Helbing et al., 2007; Helbing and Mukerji, 2012) are low resolution and the visible part of the human body, mostly the head, is represented by few pixels only. In addition, the main part of the body, including feet, legs, and torso, are generally not visible due to occlusion. It is thus often impossible to identify why people stumble or fall. This problem of occlusion pertains even to recent laboratory experiments (Adrian et al., 2020; Garcimartín et al., 2018). First experimental studies dealing with this problem are restricted to few individuals (Li et al., 2021; Wang et al., 2018, 2020). Potentially, the problem of occlusion can be addressed by combining video footage with additional sensors to record three-dimensional motions (Feldmann and Adrian, 2023). Furthermore, retrospective interviewing of people who have been in dangerous crowd situations offers another methodological strategy to learn about the physical dynamics in a crowd and the effects on the body. Johnson (1987) investigated the crowd accident at "The Who" concert in 1979 by analyzing a set of approximately forty interviews that police or members of the press had conducted with attendees at the event. He reports that people lost contact with the ground due to the pressure and were carried along by the movement of the crowd and that, after the doors were opened, people fell on top of each other and formed a pile.

The present analysis of the Love Parade disaster follows directly from Johnson's (1987) work. While several research articles have used publicly available materials such as video footage to reconstruct the Love Parade disaster (Helbing and Mukerji, 2012; Klüpfel, 2014), model or simulate its dynamics (Pretorius et al., 2015; Zhao et al., 2020), or test video observation tools (Huang et al., 2015; Krausz and Baukhage, 2012) the work presented in this article uses the complete set of witness statements in order to look inside the crowd. Immediately after the disaster and during the following weeks, many witnesses were interviewed, including visitors, organizers, administrators, and law enforcement. These witness statements were evidence in a court case which ended without a verdict in 2020. Only now, more than ten years after the disaster, have they become available for scientific analysis due to the discontinuation of judicial proceedings. We use the perspective of the people in the crowd to learn what collective dynamics are experienced and exactly how they affect the body. From this we expect to gain important insights into the role that the human body plays in these dynamics.

Furthermore, including eyewitness accounts in the form of interviews, questionnaires, or testimonies makes it possible to describe how people experience a dangerous crowd, what they perceive, and how they behave. These psychological aspects of dangerous crowds have often been the subject of controversy. On the one hand, the image of "mass panic" still pervades media and scientific reports on accidents (for a critical review, see Haghani et al., 2019) as well as lay people's understanding (Lügering et al., 2023): it is assumed that people panic and behave irrationally and egocentrically in this state of emergency. On the other hand, numerous scientific articles have rejected this idea, dating back even into the 1950s (Keating, 1982; Mintz, 1951; Sime, 1980). In fact, analysis of eyewitness accounts—for example, of the London bombings (Drury, 2009), the terrorist attack on "Le Bataclan" (Dezecache et al., 2021), the Costa Concordia disaster (Bartolucci et al., 2021), or the aforementioned "The Who" concert (Johnson, 1987)—shows that people in life-threatening situations engage foremost in helping behavior, even toward strangers. Admittedly, this does not mean that people always or only help: eyewitnesses report both helping behavior and competitive behavior (Bartolucci et al., 2021; Dezecache et al., 2021). In their article, Dezecache et al. (2021) argue that, especially in dynamic hazardous situations, local constellations of risks and opportunities for action determine whether people help others or focus on their own safety. The scientific question of crowd behavior is sensitive in part because normative and political discussions of responsibility and culpability are directly tied to it. In many instances, the attendees of such events have been blamed at least to some degree—whether by the media, law enforcement agencies, or politicians—and accused of panicking or behaving aggressively. This has been intensively worked through, among others, for the Hillsborough disaster (Hillsborough) or the "The Who" concert (Johnson, 1987). In this article, we reconstruct how visitors to the Love Parade perceived the situation, what they experienced (emotionally), and how they and others behaved. In doing so, we adopt a descriptive perspective that remains close to the concrete statements of the witnesses and does not initially inquire into the causes of specific psychological processes.

To sum up, research on pedestrian and crowd dynamics in general, and reconstructions of the Love Parade specifically, draw heavily on video recordings. Therefore, crowds are usually described and analyzed from the set perspective or top view. Complementing this, this article takes the perspective of the people in the crowd. This allows us to consider aspects that cannot be seen on videos: Movements of feet and legs, physical sensations, intentions, emotions, and perceptions. Five research questions guide the analysis of the witness statements:

- What happens to the body in life-threatening dense crowds?
- What collective dynamics do eyewitnesses describe?
- What do they perceive in this situation?
- How do they behave in this situation?
- How do they experience this situation emotionally?

Before we present the results of our content analysis, we outline the events which led to the Love Parade disaster and describe the local infrastructure. This article in no way—neither in the reconstruction of the disaster nor in the analysis of the witness statements—seeks to ascribe blame for what happened.

## 2. Brief description of the Love Parade event and disaster

The Love Parade, a dance parade led by slow moving music trucks with extremely loud electronic and techno music, was first held in Berlin, Germany, in 1989. In the 1990s, the number of participants grew quickly, reaching several hundred thousand visitors. As a result of the sheer number of participants and problems with the garbage and waste they left behind, the Love Parade was publicly criticized. When authorities in Berlin refused permission for the event, it moved to the Ruhr region. After events in Essen, Dortmund, and a cancelled event in Bochum, the Love Parade was held in a former freight station in Duisburg in 2010. As guests were admitted through the entrances, there were repeated lengthy periods of congestion. In the afternoon, people filled the main ramp leading to the event site, eventually forming a very dense crowd in which five hundred visitors were injured and twenty-one died.

**2.1 Access to the event and creation of congestion on the ramp east**

This chapter summarizes individual incidents leading to the deadly congestion at the ramp between 4 and 5 p.m. on the day of the event. An overview of the venue and the access routes can be found in Fig. 1. All of the information is taken from the final report submitted by the expert witness for the court, Gerlach, in May 2021 (Gerlach, 2021). The description here is limited to the issue of access to the event site. It addresses when and where the flow of visitors became congested and how the responsible security forces and the police reacted to resolve these problems. It focuses on the main incidents and actions. Those interested in further details and additional information are encouraged to refer to the more comprehensive report.

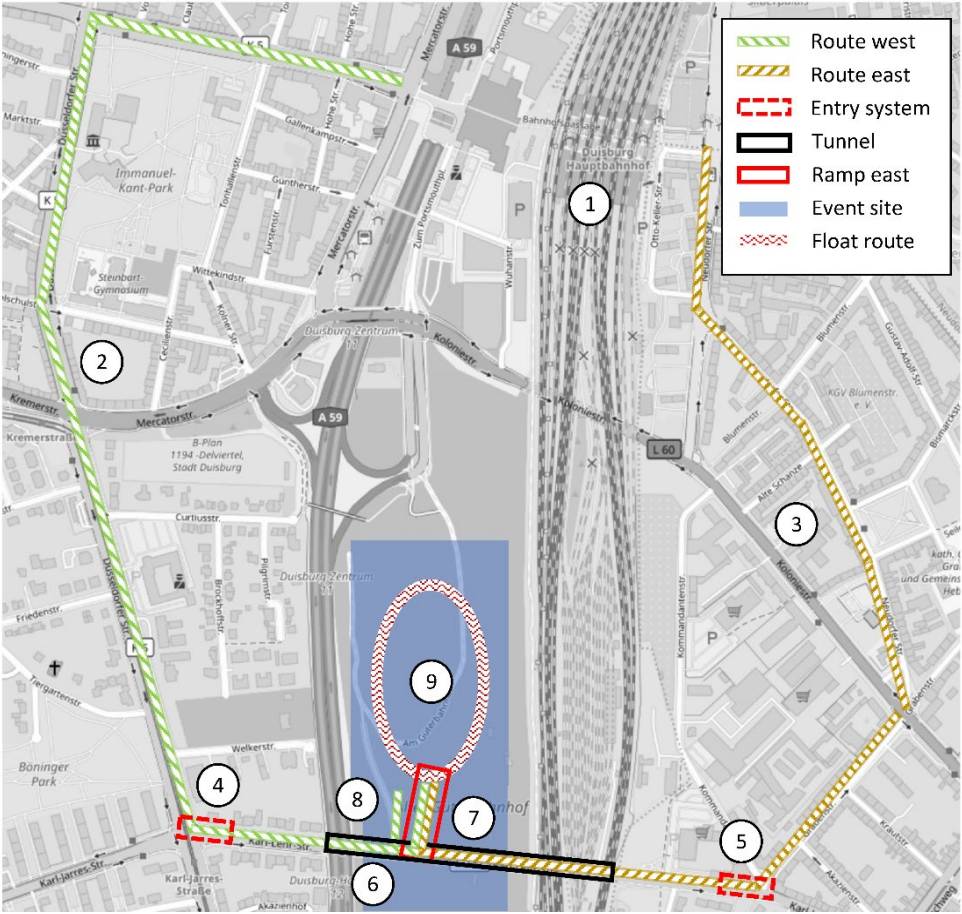

**Figure 1:** Access routes to the event site: From the train station (1) two routes led to the venue. Both the route to the west (2) and to the east (3) led to one of the respective entrance gates (4 and 5), which were constructed of line-up gates and temporary fences. Past the entrances, the routes headed to the tunnel Karl Lehr-Straße (6). From the tunnel, two ramps (7 and 8) led to the site of the event (truck route) (9). But the ramp west was intended as an exit only (8). The deadly crush occurred at the ramp east (7).

The majority of visitors arrived via the train station and walked from the train station to the venue via two routes, one to the west and the other to the east of the venue. The routes lead to a tunnel from which two ramps (ramp east and ramp west) led to the event area. In order to regulate the flow of visitors into the tunnel and close it off if necessary, two entrance systems constructed of line-up gates and temporary fences were installed on Karl-Lehr-Strasse at intersections in front of the western and eastern tunnel entrances. The temporary fences around the line-up gates also separated the incoming and outgoing streams of visitors and were to be patrolled by security personnel. The organizers expected that visitors who had reached the event area via the ramp east would naturally follow along behind the passing trucks. The event area was designed as a circular course along which the music trucks drove continuously. The following description lists the main incidents and those measures that directly influenced the flow of visitors and the congestions that occurred at different points along these routes. This article omits all other aspects and problems, like communication between police and security personal and the lack of possibilities to communicate with the visitors and others present.

The event's start, scheduled for 11 a.m., was delayed by about one hour. By the time the entrance gates west and east opened shortly after noon, many visitors had already gathered in front of the entrances. Over the course of the day, significant traffic jams of arriving visitors formed in front of these entrance systems repeatedly. Critical situations occurred there again and again.

**2.2 Head of the ramp**

Shortly after 2 p.m., the float parade started in the event area. There were an estimated 37,000 people in the event area. Around 2:15 p.m., congestion was first observed in the transition area between the east ramp and the event area. Since the ramp west was intended to be exit-only, the visitor attractions could only be reached via the east ramp. According to the planning documents, the free space for leaving the head of the ramp by following the trucks, the transition area, was at least forty-five meters wide. On the day of the event, however, the width allowed by temporary fence elements was only twenty-eight meters, see Figure 10 and 11 in (Gerlach, 2021). This bottleneck was a crucial factor for the development of the congestion at the head of the ramp and has not been considered in earlier works (Helbing and Mukerji, 2012; Klüpfel, 2014; Pretorius et al., 2015; Zhao et al., 2020). At this spot there was a complex cross traffic of visitors coming and going in addition to the music trucks, which passed in the immediate vicinity and which the visitors followed. Moreover, visitors got their initial view of the parade of music trucks when reaching the top of the ramp in the event area. Some arriving visitors paused at this point to observe the scene. All these factors led the crowd to grow in this transition area between the ramps and the event area as the event continued after 2:15 p.m. Later on, this crowd tailed back to extend into the ramp east. The multiple factors at the top of the ramp contributed to the congestion at the event site and had the potential to lead traffic to a standstill on the full ramp and in the tunnels.

From 2:20 p.m. onwards, the music trucks had to stop again and again due to the crowds in the transition area. They were therefore unable to help clear the congestion by encouraging visitors to move along away from the ramp. Additionally, an increasing number of visitors sought to leave the site (earlier than expected), using the ramp east rather than the ramp west as planned. This created a bidirectional flow on the east ramp. A number of measures were initiated to solve the problem of the crowd backing up at the ramp. Security personnel were instructed to act as pushers to encourage visitors at the head of the ramp to go along with the music trucks. In order to reduce the congestion in front of the entrance system west, additional barriers were set up by the police on the route west and visitors were redirected to the route east. Due to the redirection, the flow at the route east increased. At the entrance system east, the congestion increased and the situation later also became critical. Further barriers were also installed along the route east between the train station and the entrance system.

Around 3:30 p.m., the tailback at the top of the east ramp had grown to such an extent that fences on both the west and east banks along the side of the east ramp had been thrown over and overrun. The police and the organizer's crowd manager decided to temporarily close the two entrance systems west and east in front of the tunnel, set up a cordon of police forces on the ramp east, and open the ramp west for the inflow.

**2.3 Entrance systems west and east**

At approximately 3:55 p.m., the entrance systems east and west were temporarily closed. The closing of the line-up gates led to congestion and crowding in front of the gates, which were also critical. There was a risk that the temporary fences would fail due to the pressure. Although law enforcement officers had ordered the closing of the entrances, the entrance system west was kept open continuously from 4:02 p.m. to 4:55 p.m., and the entrance system east was open and closed intermittently during this period. At the entrance system west, fence elements were opened at around 4:31 p.m. to relieve the congestion and clear a path for a rescue vehicle. At 4:35 p.m., the opening in the fence was expanded to relieve the congestion. The entrance system west was completely abandoned, and all those persons standing in front of the entrance system were able to flow into the west tunnel unimpeded. At 4:40 p.m., the wide-open fence at the entrance system west was closed again, and the congestion in front of it had dissipated. At 4:45 p.m., stewards occupied the passages of the entrance system west again. At 4:55 p.m., the entrance west was finally closed by police forces.

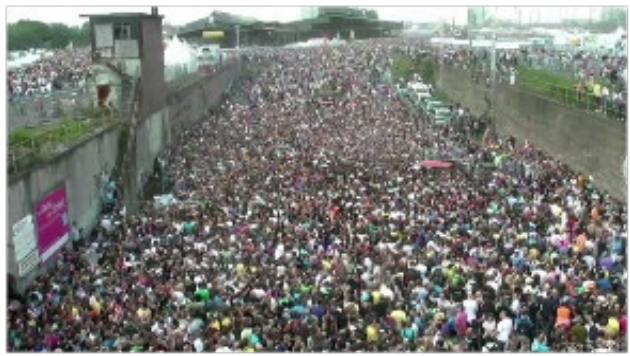 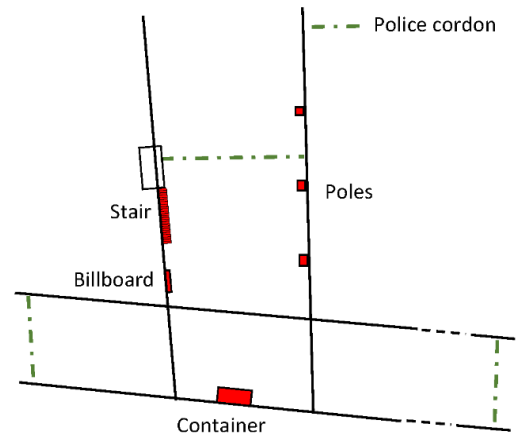

**Figure 2**: Left: Snapshot of the ramp east at 4:25 p.m., shortly after the third police cordon was dissolved (snapshot from https://www.youtube.com/watch?v=3z1Sk2OSIOw). The opposing streams were blocked at the center of the ramp east. Right: Sketch of the ramp and position of the police cordons (green dotted line). The staircase, the poles and the container were perceived by visitors as possibilities to enter the event site.

## 2.4 Police cordons

Three police cordons were set up between 3:50 p.m. and 4:00 p.m.: one in the tunnel at the west, one in the tunnel at the east, and a third cordon centrally on the east ramp. On the east ramp, the police cordon blocked the bidirectional flow (incoming visitors at the bottom, leaving visitors at the top). At the place where the third police cordon was established, temporary fence elements narrowed the ramp, so that there was little need for personnel to form the cordon. Some of these fence elements were essentially without function; they had served to prevent unauthorized access to the site prior to the event and had not been removed due to negligence. At around 4:15 p.m., the second police cordon in the east tunnel could no longer be held. At around 4:20 p.m., the police cordon in the west tunnel was likewise dissolved, and more visitors streamed onto the east ramp. At the same time, the east ramp was increasingly filled with people who wanted to leave the event. These opposing streams of visitors were blocked by the third police cordon, which was positioned centrally on the ramp. At around 4:24 p.m., the third police cordon on the ramp was also dissolved, as the police forces were pressed between the crowds of people moving in opposite directions. People stood crowded next to each other and the two opposing streams of people blocked each other. In the middle of the east ramp, there was a standstill.

## 2.5 Standstill on the ramp east

A very dense crowd developed in the lower and middle areas of the ramp. In these areas, the ramp (which emerged from the tunnels east and west) was surrounded by high walls (see Fig.2). Along the ramp there were three light poles, a narrow staircase to an abandoned signal tower, and a container for the crowd manager. Each of these was secured with temporary fences. These were torn down, and, at about 4:23 p.m., the southern light pole was used to access or climb into the event site, followed just a few minutes later by the container and the narrow staircase to the signal tower. The first people who escaped the congestion and successfully reached the site could be easily seen by the people who were stuck in the congestion on the ramp.

## 3. Method: Content analysis of witness statements

The empirical material for this article consists of written transcripts from witness interviews conducted by police officers in the hours, days, and weeks after the Love Parade disaster. A similar corpus of empirical material was also used by Johnson (1987) and Bartolucci et al. (2021) to reconstruct crowd disasters.

The witness transcripts used here vary in detail and format: Whereas some contain the questions of police officers and verbatim transcriptions of witnesses' answers, others only summarize what the witness said. In some cases, a structured questionnaire was used, in other cases, witnesses talked more freely about their experiences. In many cases, witnesses were asked to look at a map of the area in which the dense crowd was standing and locate their own position and sometimes that of others. Furthermore, some interviewees wrote about their experiences and observations themselves and sent these reports to the police. These texts were included with the witness statements. Pictures that witnesses attached to their witness statements were not included in our analysis.

In total, thirty-two file folders (paper version) of witness statements in alphabetic order exist. We selected our material based on two criteria. First, we only used statements by visitors (rather than event organizers, law enforcement officers, or city officials). Second, we selected statements from those witnesses who had been in the area of the lower east ramp (see Fig. 2, right) between approximately 4 and 5 p.m., in other words, during the period when the crowd became dangerously dense and people died or were injured. We estimate that nine hundred extant witness statements fulfill these selection criteria. Of these, we randomly selected one hundred witness statements for the initial analysis.

In order to describe and systemize the descriptions of events as related by the witnesses, we began by analyzing the qualitative content (Mayring, 2000) of the data. The categories were first developed deductively and then inductively revised and complemented. We began by structuring our categories according to the five research questions listed above. Then in a preliminary step of our analysis, we made a list of sub-categories based on existing analysis of the Love Parade disaster and similar catastrophes (Bartolucci et al., 2021; Drury et al., 2009; Helbing and Mukerji, 2012; Johnson, 1987) (table 1, black). Additional categories were added during the initial reading of the complete sample of witness statements (table 1, blue).

| Descriptions of body and impaired bodily functions | Collective dynamics | Perception within the crowd | Behavior (of oneself and others) | Emotions and emotional expressions |
|---|---|---|---|---|
| Legs/feet | Wave motion | Perception of stairs/pole | Helping behavior | Anxiety |
| Arms | Piling up | Free space is perceived on the ramp | Pushing | Aggression |
| Falling | Domino effect | Missing information | Aggressive behavior | Party atmosphere |
| Breathing problems | Hole | Change of interpretation: normal situation becomes a dangerous one | Giving up | Screaming (for help) |
| No contact with the ground | Counterflow | Body size as factor | Climbing up stairs/pole/container | Crying |
| Upper body bruised | Clogging in front of stairs | | Moving toward stairs/pole | No fear |
| Stumbling | | | Climbing over others | |
| Dead bodies | | | Focusing on oneself | |
| Fainting | | | Trying to stay with others | |
| Neck/head | | | Forming chains | |
| Nausea | | | | |
| Face color | | | | |
| Getting caught with bag/loosing bag | | | | |
| Temperature too hot | | | | |

**Table 1:** Categories for content analysis, black: deductive categories, blue: inductive categories, green: inductive categories from re-analysis

The content analysis was performed by the two authors of this paper together. Each statement was read aloud sentence by sentence, and small thematic units were separated. These units range in length from half a sentence to three sentences. We then coded the content of the units by assigning the unit to a category (table 1). Each unit was counted separately, meaning that a single category can appear twice or even several times in one witness statement (for example, when different instances of helping behavior are mentioned). We sorted the results according to their frequency to provide a relative relevance and to inform about the magnitude of mentions without aiming for a precise quantification. Furthermore, exemplary (anonymized) quotes were selected for each category.

After the completion of this first round of analysis, and after a first draft of this paper had been finished, we randomly selected another folder and selected the relevant witness statements (twenty-six additional statements). These witness statements were likewise coded. This helped us to determine whether the analysis had reached a sufficient degree of saturation, that is, whether additional analysis of further witness statements would generate additional findings. In fact, we found that the analysis of the twenty-six additional witness statements mainly added to the content of the results already described and confirmed the existing results. Only small changes in the category system were required (new subcategories were added which did not change the overall picture but allowed for more details). Two additions to the category system, however, were of a larger nature. Here, upon re-reading and re-coding, we found that our analysis became more accurate as a result of the additional witness statements. This relates first to what happened below the stairs and secondly to the perception category. The first sample of one hundred testimonies was not subjected to a new analysis with the new subcategories. Thus, our presentation of the results of the additional subcategories explicitly states which part only emerged as a result of the further step of analysis. To make the number of mentions comparable, we extrapolated the total values for the additional subcategories. The extrapolated values are indicated (number in re-analysis sample/extrapolated number in complete sample). Because the first run of re-analysis generated additional insights, we analyzed ten more randomly selected testimonies. No new subcategories emerged during this run. Therefore, we determined that further inclusion of the remaining approximately 760 witness statements was unlikely to reveal findings that contradict the results presented here. With the re-analysis, the complete sample consists of 136 witness statements.

For the purposes of scientific analysis, the credibility of the testimonies was evaluated based on criteria developed in the context of legal proceedings (Schneider, 2010). These include logical consistency, richness of detail, spatio-temporal linkage, description of gaps in memory, complications and ambiguities, self-incrimination, and questioning of one's actions. Almost all the witness statements are detailed (especially with regard to the descriptions of their own physical experiences in the crowd) and fit the general reconstructions of the accident. In numerous cases, witnesses describe gaps in their own memories (e.g., due to fainting), thus also acknowledging the limits of their own ability to testify. Some witnesses report actions for which they condemn themselves in retrospect. The fact that reports of helping behavior are dominant (see below) may possibly be related to the fact that witnesses prefer to remember these moments of humanity rather than other aspects. There is no evidence, however, that witness testimonies are systematically distorted in this respect. In individual cases, especially when the witnesses testified days or weeks after the events, one can see the influence of media coverage and suggestive questions from officials. However, some witnesses actively contradict the theories presented to them (e.g., that it was a mass panic or that people died when they tried to climb up the stairs and fell). By not relying on individual statements but incorporating a total of 136 testimonies, we eliminate the chance that individual problematic testimonies might create a false picture overall. The large amount of material also allows us to compare how witnesses describe their own behavior versus the behavior of others. In fact, both forms of reporting are consistent, and it does not seem that the witnesses try to make themselves look better by describing their own helpful behavior omitting mention of any selfish behavior. Only two witness statements were excluded from our analysis because they were inconsistent. (Their description did not resemble the events reconstructed above; presumably they were at ramp at a different time and had confused the time in their witness statements.) These were replaced by two other witness statements.

In addition to the content analysis we noted age and gender of the selected eyewitnesses as reported in the protocols.

The witness statements were made available to the authors for the period of evaluation by the public prosecutor's office in Duisburg, Germany. The authors agreed to use the data only for the academic purposes described here, to store them securely, and to publish them only in anonymized form.

During the analysis, it became clear that an interpretation of the material that went beyond the descriptive approach of content analysis would yield additional insights. Among other things, questions of sociality (Did visitors feel part of a crowd? With whom did they identify, from whom did they distance themselves?) require such an interpretative and reconstructive approach (e.g., Bohnsack et al., 2010), since the visitors do not explicitly talk about these aspects (but an interpretation, for example of the use of "we," can bring up implicit content). We decided to exclude this part of the analysis from the present article due to its complexity.

## 4. Results

First, for each category the results of the analysis are reported in descending order of frequency. Exemplary text passages of the testimonies (translated from German to English) are added to the analysis. To preserve anonymity, the text passages have been paraphrased. In addition to presenting category by category, the overall picture and some of the findings are discussed.

The average age of the selected witnesses in 2010 was 25.8 years (ranging from 16 to 53 years), 43% were women, 57% men.

### 4.1 The body in life-threatening dense crowds

Most often, **problems with breathing** (61) were mentioned. When these problems were explained in more detail, they were attributed to either pressure on the upper body (20) or poor air quality at the bottom of the crowd (21). Here, the pressure is described:

> "The reason, in my opinion, was that there was such pressure that I couldn't breathe for two or three breaths. I could not lift the chest."

> "Due to the pressure, I could not breathe for several seconds up to one minute, which caused some of my blood vessels to burst and later my body was red throughout."

> "The people in front of me were pressing so hard against my chest that I thought I was basically going to be crushed and suffocate. I suffered pain that you can't imagine."

> "My chest was being compressed more and more. I was just gasping for air and trying to push people away from my chest. At some point I managed to get my arms in front of my ribcage and protect myself a little bit."

The poor air is described as a problem particularly for shorter visitors, who tried to stretch upward to breathe:

> "It got tighter and tighter, and I stood on my tiptoes to get better air."

> "It was worse at the bottom than at the top. I had to stretch my head up to get air."

The second most frequent mention was **fainting** (60). Both others' fainting and the witness's own fainting are described (some of it over and over again).

> "I saw at that time around me, I could overlook a crowd of maybe one hundred people, several of whom I assumed were unconscious. So I'm not sure, but the heads were sometimes hanging down crookedly and others were shaking and screaming and somewhere I also saw that someone was trying to give air to someone who was squeezed in standing up and somehow took air at the top and blew in at the bottom."

Often, this is associated with the first category, problems with breathing. Visitors repeatedly describe how people fainted because of the lack of oxygen.

> "The air became less and less. At some point, all I saw were black dots in front of my eyes. Then I passed out and couldn't see anything."

Most people who fainted also fell down. Some were held up by others.

> "It didn't take long before the man fainted again. I noticed myself that it was getting tighter and tighter. I tried as long as possible to hold the man, but then it was no longer possible. I had to protect myself. The man slowly slipped away downward. Then I didn't see him anymore, the people were immediately pressed together above him. There was simply no one who could get out of the way."

**Falling** is described thirty-two times in total – with or without fainting as the mentioned cause. These cases only encompass those instances in which a single person fell. The phenomenon of several people falling at once is described in the category "collective dynamics" below. Also, quite frequent are descriptions of **problems with the legs and feet** (32). This includes someone standing on a foot or leg so that it cannot be moved anymore (10), losing a shoe because it got stuck (14), a leg being trapped (6), no room to put down a foot (1), and fear that a leg could have been broken (1). There were also reports of difficulties staying on one's feet. Some people describe **stumbling** (9), others state that they did not have **contact with the ground** anymore (9). They describe how they stumbled either over objects on the ground or people who had fallen on the ground themselves:

> "Next to me, a young man stumbled over a person lying on the ground. But one could not see, even in the immediate vicinity, where and whether people were lying on the ground. A man next to me kept saying that his girlfriend was underneath, but you could neither see nor feel anyone."

With medium frequency, **heat** is mentioned (13) as well as **problems with the head or neck** (14). Some visitors report that others climbed over them and pressed on their heads.

Furthermore, it is mentioned that the arms and hands of others are pressed into the neck or face. This corresponds to reported **problems with witnesses' own arms** (14). Visitors describe that they no longer had control over their arms and that they pressed others:

> "I didn't have my arms on my body, but I also didn't have control over them anymore. I couldn't have brought them back to me. My left hand was in a girl's face. She then bit my hand to make it clear she couldn't breathe.'

One woman describes that someone grabbed her hand so tightly that she had sores afterward.

Several witnesses describe seeing people who were **dead** (18) lying on the ground, or who they suspected were dead. This is often accompanied with a description of the **facial appearance** (5) being pale, blue, or bloody.

A few mentioned that their **bag got caught** (4) and that they therefore tried to get rid of their bag, and others tell how their bags were torn away (2). Three mentioned that their **upper body was pressed against the wall**. Two witnesses mention **nausea**. One person was hit by a rope that others had used to climb out of the crowd.

Very clearly, breathing is a serious problem in very dense crowds. The loss of the upright position is furthermore a frequently described problem. Several categories can be summarized here, namely, falling (32), stumbling (9), no contact with the ground (9), and other problems with legs and feet (32). This confirms the assumption that falls are a dangerous phenomenon not only in fast moving crowds (like a stampede) but also in slow moving, very dense crowds. In the case of the Love Parade there was some discussion about whether people stumbled and fell over objects on the ground, both in the media and in the questions police officers asked in the interviews used in this study. The data does not confirm this hypothesis; only four witnesses mention stumbling over objects on the ground, and two other witnesses explicitly stated that objects on the ground were not dangerous for them. Stumbling over other people, however, was mentioned and also played a large role in the collective dynamics described below. So if they did not stumble over objects then why did the visitors fall? From the data we can derive that fainting might have been one of the reasons. As far as we know, fainting has not been in the focus of crowd research very much. In particular, experimental research with crowds is always conducted with fit individuals who participate in an experiment for a span of minutes – but not hours like some who tried to enter the Love Parade. If we imagine a dense crowd of only strong and fit participants, it is much harder to imagine why people fall than in a crowd of participants with circulatory problems. If people get weak, they fall easily; if they faint, they can only remain upright with the help of others. It can be concluded that, in addition to acute dynamics in a crowd, factors such as time (how long have people been standing in the crowd), climate (how hot is it? what is the air flow like?), nutrition (when was the last time people were able to eat and drink?) can be crucial.

### 4.2 Collective dynamics

Quite often, **wave motions** in the crowd are described (55).

> "You also have to imagine that the crowd got worse and worse in waves and that you lost your upright position more and more with each wave and your upper body was pushed forward. In principle, I did not move my feet."

Whereas some descriptions indicate whether the movement was from front to back (like in the last quote) or left to right, most witness statements remain vague in this regard. Some participants describe how they tried to deal with the waves by going with the wave without resisting it or by absorbing the pressure and keeping their balance:

> "Then came these pressure waves. I felt pressure from behind and had to try to compensate with my upper body so as not to fall on other people. You could no longer move your feet, because there was no room to the side. These pressure waves were actually always there, sometimes more, sometimes less. At some point it was so tight that you couldn't compensate for anything with your upper body."

This quote makes clear that compensating for the loss of balance caused by the pressure requires space for the feet and the upper body and therefore becomes more difficult the denser it gets. Other witnesses describe feeling completely helpless and report many cases of falls (their own or others).

> "All of a sudden, I just noticed that the pressure behind me went away, and I myself fell backward toward the wall. I fell onto the people who were standing behind me."

In this quote the fall is associated with a sudden change of direction of the wave. In other cases, it sounds like the waves pushed people down:

> "Once a wave came that somehow pushed me to the ground. I fell to my knees and could still support myself with my hands. People from behind stepped on my legs. Some fell on top of me."

It is furthermore mentioned that people had to walk or climb over others who were lying on the ground because they could not resist the wave (17).

> "Due to the fact that I probably completely lost my upright position at some point, I was pushed to the ground, but I managed to at least still push my upper body up with my arms. Since I did not manage to stand up, it was then that some people stood on top of me during the next wave. To this I must say that this situation was of course much more than only frightening for me. Since I didn't know any other way to help myself, I pinched the leg of a girl in front of me and told her that if she didn't make sure that I was pulled up, that I wouldn't be able to take it much longer. This girl then told all the people standing around me and somehow they all managed to pull me up together."

"When I stood again I saw in the immediate vicinity, I would say one or two meters away from me, a young woman lying on the ground. I shouted that she should be helped and could see that she had blood on her face. She had one eye closed and the other open. Shortly after, I could no longer see her, because the crowd simply trampled over her."

A total of thirty-four statements described how people fell over each other, **piling up**. Two witnesses speak in such a case of a "domino effect."

"Then it became so tight that you were pressed from the front and it became like a wave movement, always from the back to the front and back. Through this wave movement, this whole pile then fell over at some point."

Some witnesses describe this falling as a particularly slow movement:

"You fell to the ground because you were carried away by the others who were falling. Even if you were standing safely, you fell because a large crowd was leaning on you. You didn't fall quickly, but rather you tilted very slowly toward the ground until you got there."

In some witness statements the piling up is attributed to a "**hole**" in the crowd (7).

"Here I noticed that the first people went down. Why they suddenly sank away, I don't know. They went down on their knees and a hole formed. The people standing next to it tried to keep themselves from falling over the kneeling person by supporting each other. But eventually the strength wore off, and they were pushed into the gap. The kneeling person was then no longer visible."

"I still saw someone pass out and slump down. Some others wanted to pick him up again and bent down. But because the crowd was pushing from behind, these people also fell down. And then more and more tripped over them and then there were many people lying on top of each other. It was a whole mountain. There must have been at least thirty-five people."

"When a gap is created, you could say that the people all around are pressed into this gap and thus fall. The people in the second and further rows inevitably fall into this gap due to the enormous pressure."

People lying on top of each other were so wedged in that they could not get out of the situation, at the same time more and more people fell on them. It is described that more than eight people were lying on top of each other and a total of between thirty and forty people were stuck.

"The people have more or less wedged themselves together in the fall and lay wildly, sometimes four or five on top of each other."

From the reports, it can be seen that this pile hardly dissolved, even when it was less dense around it. As a result, people lay in this pile for a relatively long time.

In the second phase of the analysis, it became clear that two different occasions of people piling up can be distinguished, one can probably be localized in front of the billboard advertisement and one directly at the bottom of the stairs (see Fig. 2, right). At the bottom of the stairs, there were also more reports of **clogging**, and the authors assume that the clogging developed into the piling up. Witnesses describe how people got stuck at the bottom of the stairs and became wedged.

> "However, the chaos in front of the stairs then grew. The pressure increased and people seemed to be squeezed in and started shouting. Then a policeman, maybe two, came down the stairs. They then wanted to pull the people out of the wedged mass of people."

Finally, in eight witness statements, a **counterflow** is described. Also, witnesses describe that a police bus drove into the crowd, making the situation worse (5).

From the descriptions of these collective dynamics, it may be possible to reconstruct what caused injuries and deaths in the crowd: It is always about the combination of falls and wave movements. From the testimonies, four different reasons why people fall in the wave movements can be identified, namely, 1) unconsciousness, 2) the inability to maintain balance due to a lack of space for the feet or upper body, 3) the presence of other people already lying on the ground, and 4) sudden changes in the direction of the wave movement. It becomes particularly dangerous when several people have already fallen, creating a "hole" in the crowd into which others are inevitably pushed because there is no possibility to build up counterpressure. It is important to note that these life-threatening dynamics are rather slow to develop, rather than fast and might thereby also be difficult to see from the outside.

**4.3 Perception from within the crowd**

Almost all of the witnesses describe **seeing other people climbing the poles or stairs** out of the crowd (113). Some (29) perceived this as a way to get onto the festival grounds, others (23) viewed it explicitly as an escape route. At this point we must once again emphasize the spatial conditions: Since the crowd was surrounded by walls, people climbed out of the crowd at three points (poles, stairs, containers). Of essential importance was that, although the density of the crowd restricted witnesses' perception of the environment, these events were visible to many people in the crowd because they took place above the peoples' heads.

The following results have been found during the re-analysis phase of 36 witness statements, in which the authors realized that some perceptions had been overlooked in the first round of analysis of 100 witness statements. Some people (3/11 – three statement were found in the 36 witness statements. We extrapolated this number to the complete sample of 136 witness statement resulting in eleven findings) noticed at a certain point that, although it was still extremely crowded around the stairs, the poles, and the container, more **spaces opened up in the middle and upper areas of the ramp**. Most of the visitors, however, did not notice these opportunities because they had become focused on climbing out.

> "To get up there, I thought, I would actually only have to go up the ramp. Funnily enough, because the people from above didn't push so hard, I just walked up the ramp [...]. I then went in the direction of the stairs, that is, where the stairs ended. There I then photographed how the people were cared for. [...] So it took me about seven minutes, from the foot of the stairs, up the ramp and to the top. I was still thinking, why doesn't everybody go this way?"

**Body size** was seen as an important factor (4/15), people report being tall as an advantage because they could see above others. Short people were more likely to become disoriented:

"I am so small at 161 cm that I could not see at all over the heads of the crowd. Therefore, I don't know whether we finally got to the left or to the right on the ramp."

Some visitors complained that they **did not see any signs and that law enforcement officials did not know the way or made misleading gestures** (3/11):

"One of the policemen then made hand movements, as if he would now ask everyone to come to the stairs. At least that was the impression. One could not recognize whether he meant individual or certain persons. He waved the people toward him with both arms."

In particular, visitors who wanted to leave the festival area and walked down the ramp were confused because this seemed to be the only exit. In addition, the crowd did not appear as dangerous from the top of the ramp. This is why some visitors walked into the dangerously dense crowd from the direction of the festival area where it was not so crowded:

"Question: You see the terrible crowd on the ramp and you see how people are pulled out via stairs and containers and then you go into this crowd? Why did you go into the middle of the danger?

Answer: I wanted to get off the site. There was nowhere to get out. There was no other exit, there were only fences everywhere. You couldn't get through anywhere, the only way I could see was down the ramp."

Some witnesses (3/11) describe that the **situation appeared to be normal for quite some time** because large crowds are to be expected at festivals. The following quote shows this process of realizing danger:

"We didn't even know at first that the stairs were actually blocked off. We were initially surprised that it went so slowly and that people had to be helped up. You could also see some desperate expressions on the faces of the stewards. To be honest, I sometimes thought it couldn't be true. It's not easy when someone resists, but it took them about 1.5 minutes with three people to pull up someone who wasn't really that tall. […] We were still maybe two, three meters away from this staircase, where you noticeably couldn't make any more movement toward it, and that's when the first girl was pulled up the stairs, unconscious, by her feet. And that was when it started to get really bad. That's when the first ones really started to panic. […] Probably those behind us did not know what was going on in front and did not understand the signs and thought, let's just push, as is the case at such events."

This quote is important because it shows that even people who were in the middle of the dangerous situation were late in understanding the seriousness of the situation. Important points of orientation were the facial expressions of the stewards on the stairs (which could be seen, while the facial expressions of most of the other visitors were, after all, directed forward and not visible), the sight of people who had become faint, and the perception of a panicked atmosphere. The visitor assumes that people behind him, however, could not understand how dangerous it was in front.

These results underline how slowly information spreads in a dense crowd and remains within a fairly short range. Contrary to the image of a quickly spreading panic, people realize danger in a crowd quite late. Visual information is very limited (and often people cannot see others' facial expressions), in particular for short people, who can only see those people who are standing directly next to them. When people orient themselves toward a particular goal, they might not realize at all what is happening behind them—for example, that free space has opened up. In this situation of relative disorientation, events which are visible or audible to everyone—in this case, those climbing out of the crowd via the poles or people being lifted up above the heads of others—can become enormously influential. Furthermore, many witnesses report that people screaming made them aware of the dangerous situation. Sound might actually travel better in a crowd than visual information. However, in a crowd as large as the one investigated here, the perception of individual voices is locally restricted.

### 4.4 Behavior of visitors in the crowd

By far the most frequently mentioned category is **helping behavior** (140). We tried to identify in the quotes the factors that moved people to help. First, they helped those with whom they were attending the Love Parade. But strangers were also helped. The decisive factor here seems to be the spatial proximity and the immediate perception of the need of the other. Different kinds of helping were reported, but mainly (99) lifting others up, carrying or passing them forward, and lifting them up the stairs. Those who are helped are described as very weakened and some had fainted. This helping behavior was often a joint action, meaning that surrounding people had to communicate and cooperate, for example when people were passed from one to another or lifted up:

> "The young woman next to me became unconscious in the meantime and several people, including me, noticed that she no longer had a pulse. We somehow managed to lift her up to hand her over to the rescuers."

> "This girl then told all the people standing around me, and somehow they all managed to pull me up together."

But there are also individuals who hold up others who can no longer stand on their own.

> "I must have had person A in my arms, unconscious, for 10 to 20 minutes."

Persons lying on the floor were protected by bystanders as much as possible under the circumstances (12). Some people spoke encouraging words and urged others to persevere (18). People who had fainted were slapped by others trying to wake them up (2/8) and kept in an upright position (9). Water was distributed (20), and there were attempts to improve the air supply (3). There are also cases, however, in which people asked for help (9), but others, despite their willingness, were not able to (13). People helped others even when their own lives were in extreme danger:

> "One of them, who probably died herself, saved me from death when I could not move and people climbed over us. If she had not held her arm over my head at the last moment, my fate would probably have been a broken neck."

In the following quote the witness describes how his ability to help vanished:

> "She was then gone [fallen to the ground], and I myself had no possibility to think about other people anymore. The air was literally pressed out of my body. My field of vision narrowed more and more."

As mentioned above, many witnesses spoke of visitors who **climbed the stairs, a pole or the container**, either they themselves or others whom they observed (113). That climbing up is mentioned so often is also related to the fact that it was above the crowd and thus visible to very many people. In several witness statements, someone who wanted to walk down the stairs was violently pushed up the stairs by the policemen, but this seems to be just one particular person who was apparently trying to get back to his partner. Accordingly, the **attempts to get to the stairs, the poles, or the container** also play a central role in the witness statements (85). Here, however, a differentiated view is necessary, because this does not mean that all these people pushed with all their might in this direction. Rather, most of them describe seeing people climbing a pole or stairs out of the crowds. The movement toward the stairs or the pole is then described as a combination of the witness's own intention and collective motion, and therefore included active and passive movement:

> "That's where we wanted to go, so that we could save ourselves from the crowd. We were virtually pushed there to the stairs, because all at once actually everyone wanted to go there."

Some even say that they were pushed toward the stairs unintentionally or against their will (5/18). Such statements can be distinguished from the descriptions that involve a very strong movement toward the stairs or the pole, even at the expense of others: **Climbing or walking over others who are standing in an upright position** is mentioned in twenty cases.

> "Then people started climbing over me and the others. I just thought now get out of here or you will die. I decided to take off my shoes because I was being held from below. I pulled my legs out to go over the people. Shoulder over shoulder. They called me names."

The fact that people were climbing over others, putting them in great danger, generated outrage among many. That is why this quote is so interesting, in which the witness herself admits to having climbed over others. Here it is shown that she did this because she thought she would die otherwise. One witness seems to assume that people began climbing over others after seeing people being passed overhead to the front:

> "At some point they started passing on unconscious people over the heads [of the crowd] to the front in the direction of the stairs. Then it started that people consciously actively

> worked their way forward over the heads of others, i.e., practically walked over their shoulders."

All occasions of people climbing over others can be localized in the proximity of the stairs. This is the area in which the crowd became extremely dangerous. Presumably those who climbed up were fearing for their lives *and* were under the impression that if they tried really hard they could save themselves. People further away from the stairs seem to not have engaged in such behavior.

The witness statements also show a strong need for closeness. Many wanted to **stay close to their friends** at all costs (47). Of these forty-seven, however, thirty-six report that they were involuntarily separated from the others. Only once is it mentioned that the group separated on purpose, thinking it would be easier to get out of the situation individually.

In some, but not many cases, **aggressive behavior** such as lashing out and biting is described (20).

> "One man saw that next to him his girlfriend fainted and that of course instills fear and the first thing you do then is to try to make room for her somehow. He then of course flipped out and lashed out."

But other accounts explicitly deny that aggressive behavior played a role:

> "I said it before, what was in the media about the ravers and a brutal panic, we didn't see that in our area. On the ramp it was such that, despite the panic of everyone trying to get to the front and out, I did not see anyone hitting or kicking anyone or anything else. So it was not that people were fighting each other, but everyone was just trying to get to the front or out."

Surprisingly, active **pushing** was only mentioned in sixteen cases, but this may well be because it seemed too obvious. In fact, almost all of the witnesses described the whole situation as a "crush" (*Gedränge* in German), which implies pushing. Less frequently described behavior included **giving up** (3), **focusing on oneself**, and **forming chains** (1).

Two results in this paragraph on behavior stand out: The first is the omnipresence of helping behavior in the crowd. This confirms other studies that have shown that helping behavior is frequent in life-threatening situations (Dezecache et al., 2021; Drury et al., 2009). It seems that the need to turn toward others and to care for them is large in dangerous situations. The second is the motive to reach the poles and the stairs as ways out. Individuals more often indicated wanting to get to the stairs than to a pole. This also makes sense in that it is physically much more difficult to climb a pole. Therefore, the members of the crowd oriented themselves in opposing directions but with a stronger trend toward the stairs. We assume that standing for a long time in such a jammed situation and the fact that nearly everyone in the crowd could see the visitors who were able to climb out created a strong motive to move toward these possible exits. This dynamic led to a densification of the crowd. However, it is not possible to deduce from the witness statements how many people pushed toward the way out and how strongly they pushed. Pushing itself was not mentioned very often. But it was described that people climbed over others in order to reach the stairs, indicating a very high motivation and the willingness to forge ahead at the expense of others.

**4.5 Emotional experiences**

Emotions from the spectrum of **anxiety** such as fear, panic, or fear of death are described most frequently (78), but others reported that they were not anxious or panic-stricken (4/15). It fits that fifty-four cases mention having heared **cries or cries for help**.

It becomes clear that the witnesses have a differentiated view of the fear they experienced: Even though the term panic is used most often (51), they still do not simply reproduce the idea of a violent and selfish escape movement. The term panic is used by the majority to make clear how great the fear for one's own life was:

> "A strange mood arose in the crowd, which had something of a panic. Also in me fear rose gradually, but a kind of fear that I had not previously experienced: I was afraid of not getting out the tunnel alive."

> "Seeing myself completely exposed to this movement, struggling for air, space, and above all coolness, and seeing myself and others lying crushed on the ground, I also was filled with an incredible panic, fear, and finally mortal fear. I admonished the others in my immediate vicinity to remain calm while we sloshed back and forth, and I was probably trying to calm and admonish myself, too."

First and foremost, these visitors describe their intense fear as a reaction to a life-threatening situation. The second witness however also seems to be aware of the idea of mass panic. Probably this is why he told others and himself to remain calm. Other witnesses explicitly reject the concept of mass panic which circulated in the media in the days and weeks after the disaster. They argued that the behavior of people in the crowd was not responsible for the fatalities. A witness statement already quoted above says that a "brutal panic" did not exist and although people were panicky (afraid of dying) and wanted to get out, they did not turn against each other. The following quote clearly illustrates how one witness denied the hypothesis of a mass panic implicit in some of the interrogators' questions:

> "Police officer: Did you have the impression that there were people in the crowd who might have overreacted out of panic and thus caused an even bigger danger?
>
> Witness: Absolutely not. The people were all panicked. I think they were all having trouble breathing, and I'm sure some even stopped breathing. There were calls for help from all sides. But I also have to say that for the most part, people were trying to calm each other down and be calm as well. There was nothing that I observed that would have indicated that the people in the crowd had increased the crushing."

Furthermore, people hear others **crying** (19), which they probably perceived as a sign of despair. An **aggressive atmosphere** is mentioned in four statements, as well as a **party atmosphere** (7).

Overall, the emotional reactions described appear to be appropriate for the situation: Since people's lives were in danger, a strong sense of fear is adequate. Aggressive feelings or party feelings were only rarely mentioned. It is also interesting that, at least for a certain period of time, the emotions of the others were perceptible through screaming and crying. The testimonies show that people use the term panic to express the strength of their own fear and the fear of others. Some explicitly reject the thesis of a mass panic.

## 5. Discussion and Conclusion

During the Love Parade disaster in 2010, there were repeated blockages and very dense crowds on the way to the event area. In the late afternoon, a life-threatening situation with about two thousand visitors occurred in the area of the ramp east. Twenty-one visitors lost their lives, several hundred were injured. Approximately nine hundred visitors were heard as witnesses who were in this area at the time of the accident. The goal of the analysis presented here is to describe in a systematic and methodically controlled way what happens to people in a life-threatening dense crowd, how they perceive and experience it, what behavior they exhibit, and which collective dynamics they describe. While there are numerous analyses of dense crowds from the top view based on field studies and experiments, little is known so far about what happens in the crowd—in part because this is generally not visible on video recordings. Individual studies have used interviews or witness statements to reconstruct how people behaved in accidents (e.g., Dezecache et al., 2021; Johnson, 1987), but only one of them has dealt with high density situations in particular (Johnson, 1987).

The following key findings emerged from our analysis:

1. Helping is the behavior most frequently mentioned. Some helping involves many people who cooperate, for example, by lifting others up and passing them to the front.
2. When the first visitors climbed the stairs and the poles, this was visible to almost all people in the crowd. People subsequently tried to reach these places to exit the area in this manner, as well. Some describe that they wanted to get there intentionally, but others were pushed along with the crowd. Other visitors actively moved toward the stairs, even at the cost of others, or climbed on the shoulders and heads of others. Therefore, although helping behavior is described very often, the overall picture is mixed: Prosocial and egoistic behavior happened at the same time.
3. Many witnesses report severe breathing problems associated with either pressure on the chest or lack of oxygen. In particular, short people report bad air. Furthermore, reports of fainting are frequent. In many cases, breathing problems and fainting are linked. Fainting people are at high risk of falling to the ground, only in some cases are they held up by others.
4. Falls are described very often, and they are linked to a) fainting, b) intense dynamics in the crowd, and c) stumbling over other people laying on the ground.
5. Witnesses frequently use the term "panic" to describe their own intense fear (or the fear of others). However, they do not use the concept to refer to the lay theory of "mass panic" (Lügering et al., 2023). Some even explicitly argue that people did not act in an irrational, egoistic way but rather behaved calmly, even when they feared for their lives.

6. Pressure waves are described very often and witnesses mention being hit by these waves again and again.
7. Many witnesses report two situations in which visitors fell over each other, building a pile with more than eight people lying on top of each other. After the first people were on the ground, the people standing around were pushed and fell into the resulting gap during the next wave movement. From the described locations (in front of a billboard and at the bottom of the stairs) and the witness statements, it can be deduced that in this situation many people died and were severely injured. To conclude, one possible reconstruction of dynamics with the highest risk of dying is as follows: First single people fall due to fainting, or due to being pushed in situations without the possibility of recovery, stumbling, or sudden changes in direction of the wave movement. These falls create a gap in the crowd into which the transversal wave motions push people. Then more people fall, piling up. In this situation, people are so entangled that they can hardly get out by themselves anymore.

Our analysis shows that crowd behavior is heterogeneous. Some help the people around (sometimes cooperatively with others), while others want to get themselves out of the situation, even taking the risk of harming others. This is similar to the analysis of the Costa Concordia disaster which also showed cooperative and competitive behavior (Bertolucci et al., 2021). Which psychological factors lead to one behavior or another is difficult to derive from the witness statements, in particular with the descriptive approach chosen for this paper. These could be social psychological factors such as a shared social identity (Drury et al., 2009) or sense of solidarity (Broekman et al., 2017). However, personality factors may also play a role, or the question of what someone can perceive in the specific situation. More psychological research is needed to identify the relevant precedents. In a further interpretative approach based on this content analysis (and with the same material), we plan to examine the socio-psychological aspects in more detail.

Many people clearly wanted to reach the stairs or the poles, and some even tried to climb over others. How exactly the intentional movement of many people created the transversal waves or a strong pressure cannot be determined. One new research question evolving from these observations is how pressure from individuals in a very dense crowd adds up to create a large overall pressure in the crowd.

Furthermore, witness reports haven proven to be a valuable source for getting to know more about the three-dimensionality of the body in very dense crowds. Most important here are the dynamics that lead to falls in crowds. This is, for example, the combination of a transversal wave with a situation in which people can no longer move their feet. In this case, the pressure of the pushing cannot be absorbed. It is important to note that the dangerous dynamic described is different from the image of a stampede, which assumes that people move fast, maybe run, and then fall over or step over people who have fallen before. In the Love Parade disaster, the dynamic of falls was in comparison rather slow. Witnesses describe that they were slowly but forcefully pushed over the people who were already on the ground and that they were able to remain upright for some time before they eventually could not stabilize themselves and also fell. Previous experiments have reproduced the generation of high density to some extent (Adrian et al., 2020; Garcimartín et al., 2018; Sieben et al., 2017) and, in some cases, transversal waves (Garcimartín et al., 2018) but not how the body reacts to such pushing. Initial attempts have been made to investigate the effects of pushing on the whole body (CrowdDNA; Wang and Weng, 2018), some with the use of motion-capturing systems, but more research on pushing dynamics in crowds is needed. Furthermore, current models of pedestrian dynamics usually depict the two-dimensional projection of the human body to the ground (ellipse or circle). These models can predict the emergence of congestions or high density, but not the reactions of the three-dimensional body. Therefore, they cannot predict those processes which appeared to be most dangerous in our analysis, namely falls. The development of three-dimensional models for the simulation of crowds has just started (CrowdDNA).

Fainting and breathing problems played an important role in the Love Parade disaster. Unlike participants in crowd experiments and models, the visitors in this life-threatening crowd situation were severely weakened by the time they had already spent in the crowd, probably also by the sunny weather, the lack of possibilities to drink and eat, and prolonged standing. These factors should be considered as an additional parameter in future experiments and models: How a particular experience affects the body depends strongly on how physically resilient it is at that moment. Experiments with healthy, relaxed participants probably overestimate the body's ability to absorb pressure. Furthermore, medical input is relevant to better understanding the risk of fainting in a crowd. The witness statements indicate that some suffered from a lack of oxygen, but whether there really was a shortage of oxygen within the crowd or breathing problems were due to the pressure on the chest over a long period remains unclear. While the risk of fainting has not played a prominent role in the scientific literature on crowds so far, crowd managers are well aware of it in practice and point out the need to ensure in advance that visitors are able to eat and drink or rest regularly (personal communication). At concerts (especially in front of the stage), crowd managers watch for people who have fainted and pull them out of the crowd. Furthermore, partitioning measures are used to allow crowd managers to get to visitors who need help.

Interestingly, witnesses use the term "panic" to describe their fear of death but not to support the idea of a "mass panic." However, several officers who conducted the witness interviews—and, according to witnesses, journalists—confronted the witnesses with the narrative of a mass panic (crowds become threatening because people start acting irrationally and egoistically, hurting others in their attempts to flee). Apparently, this narrative is embedded in cultural knowledge but in this case not supported by those who actually experienced the situation themselves. It would be interesting to reconstruct everyday knowledge of crowd accidents in more detail (similar to Lügering et al., 2023).

Despite the tragedy of the Love Parade disaster, the analysis of the witness statements also shows the resilience of crowds: Witnesses frequently state that they did not fall in a wave because they could lean on someone, or that they fell, but others helped them to get back on their feet. Even people who had fainted were sometimes able to get up after they woke up or were held up by others around them. Therefore, high density in crowds should be considered both as a risk and protective factor.

**Acknowledgments**

We would like to thank the Ministry of Justice of the State of North Rhine-Westphalia, Duisburg Public Prosecutor's Office, Duisburg, Germany for making the witness statements available. Furthermore, we would like to thank Simone Tischler from the legal department of the Forschungszentrum Jülich for their support in the application process. Special thanks go to Prof. Jürgen Gerlach, who was available for discussions and gave us advice and support. In addition, we would like to thank the two anonymous reviewers for their helpful comments. The study was part of the projects CroMa and CrowdDNA. The Project CroMa (CroMa - Crowd Management in Verkehrsinfrastrukturen) was funded by the German Federal Ministry of Education and Research under grant number 13N14530 and 13N14531. Crowd DNA has received funding from the European Union's Horizon 2020 research and innovation program under grant agreement No 899739.

**Declaration of Competing Interest**

The authors declare that they have no known competing financial interests or personal relationships that could have appeared to influence the work reported in this paper.

**Data Availability**

The witness statements on which the article is based are only available through an application to the Ministry of Justice of the State of North Rhine-Westphalia, Duisburg Public Prosecutor's Office, Koloniestraße 72, 47057 Duisburg, Germany.

**CRediT authorship contribution statement**

Anna Sieben and Armin Seyfried: Writing – review & editing, Writing – original draft, Visualization, Validation, Resources, Project administration, Methodology, Investigation, Funding acquisition, Formal analysis, Conceptualization.